\documentclass[acmsmall,screen,
nonacm=true,authorversion=true %
]{acmart}

\AtBeginDocument{%
  \providecommand\BibTeX{{%
    \normalfont B\kern-0.5em{\scshape i\kern-0.25em b}\kern-0.8em\TeX}}}

\usepackage[position=bottom]{subcaption}
\usepackage{booktabs}
\usepackage{tabularray}
\UseTblrLibrary{booktabs}
\DefTblrTemplate{firsthead,middlehead,lasthead}{default}{
}
\DefTblrTemplate{lastfoot}{default}{
  \UseTblrTemplate{note}{default}
  \UseTblrTemplate{remark}{default}
  \UseTblrTemplate{caption}{default}
}

\usepackage[ruled, vlined, linesnumbered, commentsnumbered, longend]{algorithm2e}
\usepackage{cleveref}
\usepackage{xspace}
\usepackage{multicol}
\usepackage{multirow}
\usepackage{makecell}
\usepackage{xcolor}
\usepackage{footnote}
\makesavenoteenv{tabular}
\makesavenoteenv{table}

\usepackage{pifont}
\definecolor{ygreen}{rgb}{0.3, 0.8, 0.1}
\newcommand{\Y}{\color{ygreen}{\ding{51}}}%
\newcommand{\X}{\ding{55}}%

\usepackage{tikz}
\usetikzlibrary{calc,fit,shapes}

\newcommand{\titlename}{Productively Deploying Emerging Models on Emerging Platforms:
A Top-Down Approach for Testing and Debugging
}

\definecolor{jwgreen}{rgb}{0.35, 0.71, 0.1}

\definecolor{coolpurple}{rgb}{0.721, 0.141, 1}
\newcommand{\rebuttal}[1]{{#1}}

\newcommand{\parabf}[1]{\noindent \textbf{#1}}

\newcommand{\Comment}[1]{}

\newcommand{\eg}{\emph{e.g.,}\xspace}
\newcommand{\ie}{\emph{i.e.,}\xspace}

\usepackage[normalem]{ulem}
\definecolor{applegreen}{rgb}{0.45, 0.81, 0.2}

\newcommand{\sys}{\textsc{TapML}\xspace}

\newcommand{\llmfull}{Large Language Model\xspace}
\newcommand{\llm}{LLM\xspace}

\newcommand{\numSupportModels}{105}
\newcommand{\numSupportArch}{27}
\newcommand{\numEmergingPlatforms}{5}

\begin{document}

\title{\titlename{}}

\makeatletter
\let\orig@fnsymbol\@fnsymbol
\makeatother

\makeatletter
\renewcommand*{\@fnsymbol}[1]{%
  \ensuremath{%
    \ifcase#1
      \or \star        %
      \or \dagger      %
      \or \ddagger     %
      \or \S           %
      \or \P           %
      \or \|\          %
      \else\@ctrerr
    \fi
  }%
}
\makeatother

\author{Siyuan Feng}\authornote{Equal contribution.}
\affiliation{%
  \institution{Shanghai Jiao Tong University}
  \country{China}}
\email{hzfengsy@sjtu.edu.cn}

\author{Jiawei Liu}\authornotemark[1]
\affiliation{%
  \institution{University of Illinois Urbana-Champaign}
  \country{USA}}
\email{jiawei6@illinois.edu}

\author{Ruihang Lai}
\affiliation{%
  \institution{Carnegie Mellon University}
  \country{USA}}
\email{ruihangl@andrew.cmu.edu}

\author{Charlie F. Ruan}
\affiliation{%
  \institution{Carnegie Mellon University}
  \country{USA}}
\email{cfruan@andrew.cmu.edu}

\author{Yong Yu}
\affiliation{%
  \institution{Shanghai Jiao Tong University}
  \country{China}}
\email{yyu@apex.sjtu.edu.cn}

\author{Lingming Zhang}
\affiliation{%
  \institution{University of Illinois Urbana-Champaign}
  \country{USA}}
\email{lingming@illinois.edu}

\author{Tianqi Chen}
\affiliation{%
  \institution{Carnegie Mellon University}
  \country{USA}}
\email{tqchen@cmu.edu}

\renewcommand{\shortauthors}{}

\begin{abstract}

While existing machine learning (ML) frameworks focus on established platforms, like running CUDA on server-grade GPUs, there have been growing demands to enable emerging AI applications in a broader set of scenarios,
such as running \llmfull{s} (\llm{s}) within browsers and mobile phones.
However, deploying emerging models on new platforms (such as Metal and WebGPU) presents significant software engineering challenges due to rapid model evolution and limited tooling and practices for these platforms.

Previous practice for ML model deployment often follows a bottom-up fashion,
where engineers first implement individual required operators and then put them together.
However, this traditional development approach fails to meet the productivity requirements when deploying emerging ML applications,
with the testing and debugging part as a bottleneck.
To this end, we introduce \textsc{TapML}, a top-down approach designed to streamline model deployment on diverse platforms.
While the traditional bottom-up approach requires crafting manual tests,
\textsc{TapML} automatically creates high-quality, realistic test data through operator-wise test carving.
Furthermore, \textsc{TapML} uses a migration-based strategy to gradually offload model implementation from the mature source platform to the target platform, minimizing the debugging scope of compound errors.

\textsc{TapML} has been used as the default development method in the MLC-LLM project to deploy emerging ML models.
In the past two years,
\textsc{TapML} has accelerated the deployment of \numSupportModels{} emerging models in \numSupportArch{} model architectures across \numEmergingPlatforms{} emerging platforms.
We show that \textsc{TapML} effectively boosts developer productivity while ensuring the quality of deployed models.
Furthermore, we summarize comprehensive case studies from our real-world development,
offering best practices for developing emerging ML systems.

\end{abstract}


\maketitle
\makeatletter
\let\@fnsymbol\orig@fnsymbol
\makeatother

\section{Introduction}

With the recent interest in generative AI, such as \llmfull{s} (\llm{s}),
major organizations have been releasing and deploying various powerful models.
These emerging models have been empowering various emerging applications
such as art generation, coding assistance, and autonomous agents for computer use~\cite{computeruse}. 
Meanwhile, these models of billions of parameters are computationally expensive yet can be served under various computing scenarios.
As such, the silicon industry has been continually expanding the boundaries of what is computationally possible,
by innovating their computing platforms based on efficient accelerators such as GPU, TPU~\cite{tpu}, and LPU~\cite{lpu}.

From a software development perspective, deploying machine learning (ML) models is engineering-intensive.
Because there is a huge gap in mapping high-level ML mathematics to low-level machine executables,
substantial efforts have been made to develop optimized inference engines~\cite{trt,torch,tvm} to serve ML models efficiently.
Specifically, these frameworks deploy a given trained model by overloading its computation (\eg operators and sub-graphs) using platform-optimized native kernel\footnote{\emph{Kernels} in our paper refers to low-level functions that implement the computation of an operator or fused-computation of a sub-group of operators (\eg implementing convolution plus ReLU in one function). For clarity, the subsequent discussion assumes one kernel corresponds to one operator and uses both terms interchangeably.} functions (\eg implemented in CUDA/C).
Such kernel functions can be either manually crafted in a platform-native programming language (\eg CUDA)
or automatically generated by ML compilers~\cite{mlir,triton,tvm}.
Crafting new operator kernels or building a complicated compiler to generate kernels requires extensive efforts from domain experts.
Therefore, existing ML frameworks focus on supporting and optimizing a limited set of mainstream platforms such as the CPU and CUDA platforms, leaving other emerging platforms under-supported.

The models behind these emerging AI applications do not only run on CUDA functions and NVIDIA servers.
Expanding the computing diversity in serving ML models can unlock a broader range of applications and features. 
For example,
serving \llm{s} within browsers via the WebGPU~\cite{WebGPU} platform allows for local web agents without compromising data privacy.
Likewise, the new ``Siri'' based on Apple Intelligence~\cite{apple2024intelligence} depends on the neural processing unit to drive energy-efficient on-device machine learning.
These examples highlight the growing trend of diversifying computation platforms to support next-generation models and applications.

Deploying ML models is hard,
and deploying emerging ML models on emerging platforms presents an even greater software development challenge.
Beyond implementing the corresponding operators,
significant time and energy are required for testing and debugging both individual operators and the whole end-to-end model.
To this end, our work focuses on studying and enhancing the productivity of workflows that support these emerging models.
We begin by explaining the current model deployment practices, followed by a demonstration of our solution.

\begin{figure*}
\centering
    \includegraphics[width=\linewidth]{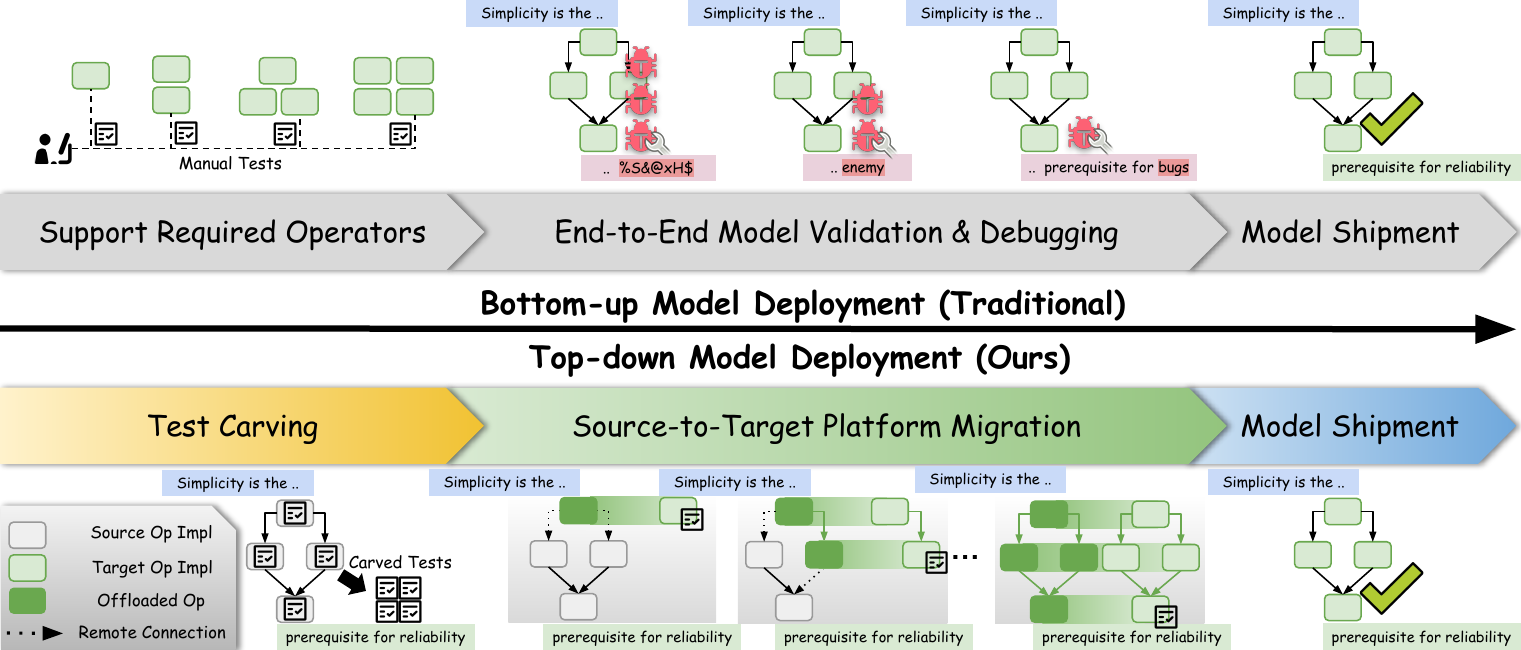}
    \caption{Bottom-up \textit{v.s.} Top-down approaches to emerging ML system development}
    \label{fig:methodcompare}
\end{figure*}

\parabf{Current practice: bottom-up deployment.}
The existing practice of model deployment naturally follows a \emph{bottom-up} pipeline, shown in the top half of \Cref{fig:methodcompare}:
\emph{(i) ``Bottom'':}
Developers analyze and implement the missing operator implementations that are required to serve the under-deployment model on the target platform efficiently; and
\emph{(ii) ``Up'':}
developers attempt to serve the model using the target-platform operator implementations and test it using end-to-end samples and benchmarks.
\rebuttal{However, such a strategy presents challenges in both testing and debugging.}
In the ``bottom'' phase, developers typically construct synthetic inputs,
resulting in test insufficiency in both quality and quantity,
due to the synthetic and unscalable nature of manual testing.
Meanwhile, deferring model-wise testing after kernel-wise testing can lead to debugging disasters:
When numerical inaccuracy occurs in the integral model testing, fault localization becomes challenging.
The error could come from inaccuracy caused by faulty single-operator implementations or propagated precision errors due to the co-functioning of multiple operators.
As such, troubleshooting in the large operator space can bring significant mental burdens, slowing down the development process.

We argue that productive deployments of emerging ML require innovations in software methodologies and toolings.
As such, we introduce \sys{}, a \emph{top-down} approach for testing and debugging the deployment of emerging ML systems.

\parabf{Our top-down \& incremental practice.}
The bottom half of \Cref{fig:methodcompare} illustrates the workflow of \sys{}.
First, we perform test carving on the model's source platform (\ie already supported mature platforms such as CUDA) by running the model with real-world inputs (\eg meaningful sentences and images).
During the model execution, we automatically instrument intermediate inputs/outputs as test cases to be used for testing operator implementations on the target platform (\ie the emerging platform to deploy the model).
Next, developers support the under-deployed model on the target platform by regarding it as a platform migration task.
The migration process starts by serving the compute graph of the model entirely on its source platform (\eg CUDA).
Each time, we implement one operator for the target platform, test it, and update the hybrid compute graph, until the whole graph entirely runs on the target platform.
Specifically, we test each target-platform operator in two steps:
\emph{(i)} test the standalone operator implementation using carved test cases;
and \emph{(ii)} test the end-to-end model via an operator offloading mechanism,
\ie replacing the corresponding source-platform implementation with target-platform implementation in the flow of computation.
Such a flow provides an error isolation mechanism where a test failure must be introduced by the operator newly implemented on the target platform.
As such, developers can quickly localize the error in a small scope and eagerly debug it while the development context is still fresh.
Additionally, emerging platforms typically lack robust developer utilities to support our methodology.
Therefore, tooling-wise \sys{} builds a universal runtime that unifies cross-platform data communication and function calling,
which can support hybrid computation graphs during migration and simplifies the testing and debugging on the emerging platform.

We summarize our main contributions as follows:

\begin{itemize}
    \item \textbf{Dimension:}
        While the optimization and abstraction for ML computation have been well studied in the past decade, we look at the productivity challenge for deploying emerging ML and address this challenge by advancing the software development approach.
    \item \textbf{Methodology:}
        We propose \sys{}, a \emph{top-down} approach to deploying emerging ML on emerging platforms.
        While the traditional \emph{bottom-up} scheme suffers from coarse-grained troubleshooting,
        \sys{} automates unit testing by carving tests from executions in mature platforms
        and adopts a migration-based approach to gradually offload the computation from the source platform to the target platform via our universal runtime.
    \item \textbf{Implementation:}
        We implemented a universal runtime to \rebuttal{facilitate} the practice of the \sys{} methodology.
        Our universal runtime currently supports \numEmergingPlatforms{} platforms.
    \item \textbf{Study:}
        \sys{} has been used as the testing and debugging method behind the MLC-LLM project~\cite{mlc-llm}. 
        Within two years of efforts, we have been applying \sys{} to deploy \numSupportModels{} emerging models on \numEmergingPlatforms{} novel platforms.
        Based on these experiences, we case study the use of \sys{} and interesting troubleshooting experiences to facilitate the future development of emerging ML systems.
\end{itemize}

\section{Challenges}\label{sec:challenge}

In this section, we discuss the main challenges in deploying emerging ML systems and why a bottom-up development methodology can be largely optimized.

\subsection{Motivating Example}

Consider the deployment of a Llama model~\cite{llama2} in a browser environment through the WebGPU platform.
ML engineers typically approach this deployment in a bottom-up fashion:

\begin{enumerate}
\item \textbf{Operator analysis:}
    The very first step is to analyze the model's operator structure.
    Models like Llama include complex operators such as embeddings, multi-head attention, root-mean-square layer normalization, and matrix multiplications.
    In reality, the required implementations can become intricate due to efficiency demands; for instance, matrix multiplication alone may require dozens of specialized variants tailored to scenarios like query-key-value (QKV) and feedforward layers.
\item \textbf{Operator implementation:}
    Once the necessary operators are identified, engineers implement their computational logic on the WebGPU platform using WebGPU Shading Language (WGSL).
    The computation logic is often inferred from academic papers, online documents, and existing implementations from other platforms.
\item \textbf{Operator-wise testing \& debugging:}
    Each operator implemented in WGSL undergoes targeted testing. %
    Engineers craft test cases according to the operator's semantics and validate the operator's accuracy by running these tests across browser platforms with WebGPU support (\eg Chrome and Safari).
    Per the testing results, they debug the operator implementation until all tests pass before moving to other operators.
\item \textbf{Model-wise testing:}
    After implementing all operators in WGSL,
    engineers attempt to reproduce full model execution in the browser by scheduling these WGSL functions according to the model's data flow.
    Specifically, for language models like Llama, developers test them by gathering prompts (\eg from open datasets or developer customization) and observing the output coherence and corresponding metrics.
    Sometimes, even when individual operators pass, model-wise validation may reveal issues\footnote{For example, 
    real-world inputs may easily trigger activations (\eg ReLU) that manual inputs cannot trigger.
    }.
    Common model-wise errors include faulty operator implementation due to insufficient operator-wise testing, accumulated inaccuracies when propagating data across multiple operators, and even platform bugs (\eg miscompilation or runtime errors),
    making debugging especially challenging.
\end{enumerate}

In the follow-up subsections, we discuss the inherent limitations of the traditional bottom-up development pipeline.

\subsection{Test Insufficiency}

\begin{figure}
\centering
    \includegraphics[width=0.85\linewidth]{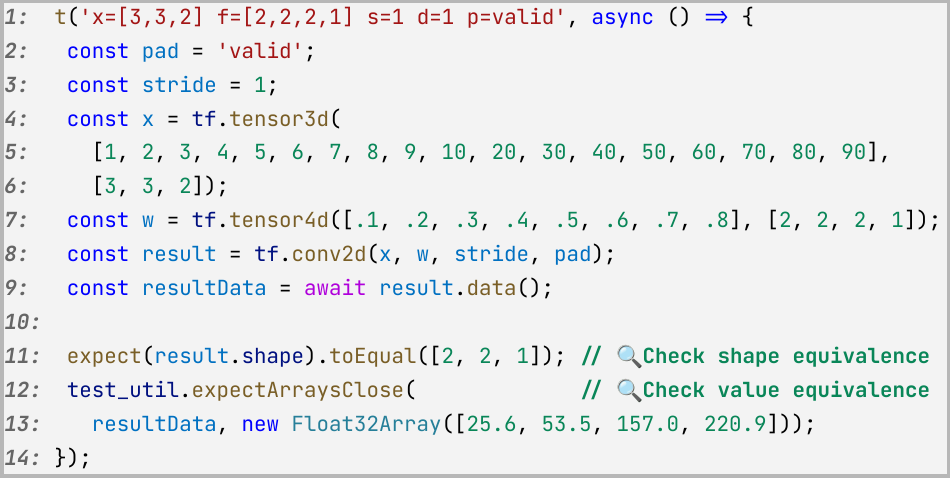}
\caption{Sample unit test in TensorFlow.js.}
\label{fig:tfjs}
\end{figure}

Complex systems such as ML frameworks require extensive testing to ensure their reliability.
For example, 40-42\% of the Python and C++ source code of PyTorch~\cite{torch} and TVM~\cite{tvm} are testing code, as of Q3 2024.
\rebuttal{While ML frameworks developed comprehensive tests to exercise implementations on mature platforms,
setting up emerging models on emerging platforms often suffers from low test coverage due to limited tests and constrained timelines.
Meanwhile, manually curating high-quality test cases is labor-intensive for developers.
For instance, the WebGPU backend of TensorFlow.js~\cite{tfjs} is implemented in 19,530 lines of code,
with only 3,549 lines of code attributed to testing.
Specifically, most testing codes are dominated by a few common operators such as matrix multiplication and convolution.
Furthermore, directly migrating tests from one platform to another is also challenging due to inconsistency in API granularity, interface compatibility, and supported data types.
}

\Cref{fig:tfjs} exemplifies a test case for \texttt{conv2d} in TensorFlow.js.
Developers craft the test case by specifying operator attributes (Lines 2-3), input data (Lines 4-7), and the oracle,
including the expected output shape (Line 11) and output values (Lines 12-13).
It is worth noting from Lines 4-7 that the test inputs are rather synthetic, given the apparent value orders in \texttt{x} and \texttt{w}.
Such simplistic test inputs may not represent the distribution of inputs derived from real-world user-given data,
whose precision deserves more attention.

\subsection{Obscure Precision Bugs}

Result inconsistency is one of the most challenging and harmful bug symptoms in software engineering.
Compared to other fail-stop bugs such as program crashes,
debugging result inconsistency is extremely hard as it only presents the error in the final result without associating the intermediate step that introduces the error.
Unfortunately, such types of errors can be common in ML due to its heavy use of arithmetic computation (\ie few operators can trigger crashes) and floating-point numbers.
For example, we found the calculation $1035 - 1031$ in \texttt{float16} erroneously leads to a result of 5 on the Qualcomm Hexagon backend.

Furthermore, despite passing operator-wise testing, result inconsistency bugs can still happen and can be even more challenging to debug in the end-to-end model execution.
This is because emerging models have increasing complexity and size and incorporate a greater number of operators of various data types.
Therefore, the cumulative impact of precision errors becomes more pronounced, and a large debugging scope needs to be checked in order to isolate the error.

Meanwhile, in the bottom-up development pipeline, operator-wise testing and model-wise testing are entirely separated into different phases.
As such, when debugging model-wise issues, it is challenging for developers to recall related operator-wise debugging contexts.
This limitation calls for applying model-wise testing while developing single operators.

\subsection{Limited Tooling in Arising Systems}

The infrastructure stack in mature platforms has been long developed,
allowing developers to seamlessly invoke device functions, transfer data between host and device, and so on.
In addition, these platforms can also provide ready-to-use debugging tools.
For example, the TensorRT~\cite{trt} ecosystem implements Polygraphy~\cite{polygraphy}, a graph reducer to localize error-inducing sub-graphs.

In contrast, emerging platforms often have limited tooling support, making it challenging for developers to analyze internal computational states—a crucial aspect of debugging. 
This lack of tools exacerbates existing challenges.
Therefore, developing infrastructures and tooling that facilitate developer interaction with these platforms is essential.

\begin{figure*}
\centering
    \includegraphics[width=\linewidth]{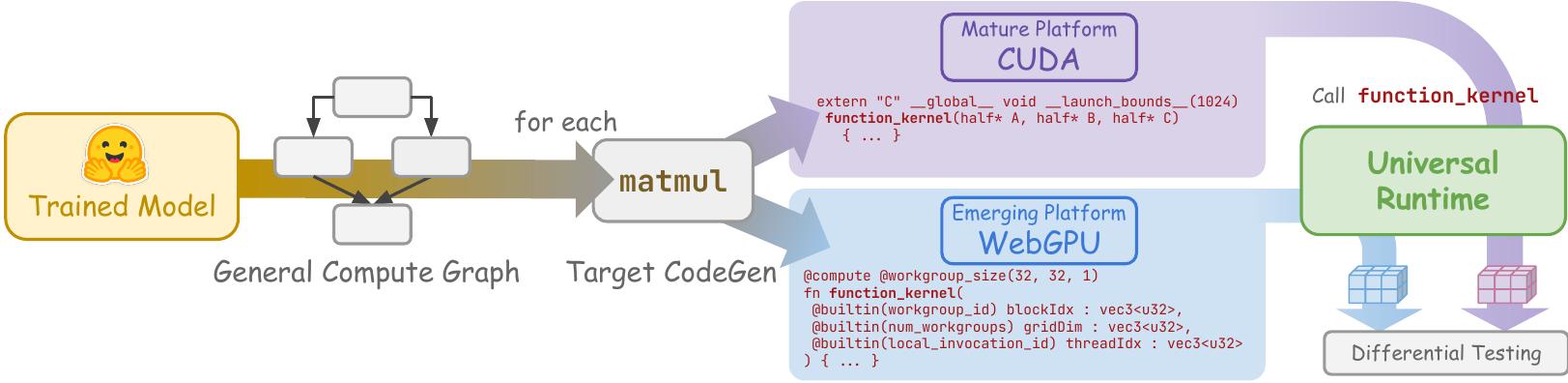}
\caption{The computation representations in the \sys{} workflow. 
To deploy a trained model, \sys{} converts it into a general compute graph, where each operator can be offloaded to various backends through target code generation.}
\label{fig:codegen}
\end{figure*}

\section{Design of \sys{}}\label{sec:design}

\subsection{Overview}\label{sec:overview}

The goal of deploying a model on a specific platform is to describe the model's computation using the native program of the target platform.
Specifically, \Cref{fig:codegen} illustrates a typical deployment workflow in \sys{}.
The input to the deployment flow is a trained model produced by the training framework (\eg PyTorch).
Subsequently, the trained model is translated by the framework into an internal intermediate representation (IR) that generally describes the semantics of the model.
Through a series of general-purpose optimizations, 
the IR flows to the target code generation stage, which transforms the computation to native programs of the target platform. 
For example, existing frameworks often specialize the code generation pipeline to CUDA, a mature platform in the NVIDIA ecosystem, and these computations are finally represented by CUDA C functions, shown in the purple box in \Cref{fig:codegen}.

We describe the goal and overall steps of \sys{} using the example of \Cref{fig:codegen}.
The goal of \sys{} is to facilitate the process of implementing and testing target code generation of the model operators on the emerging platform, \ie on WebGPU such that users can run their model on a browser, as is shown in the blue box in \Cref{fig:codegen}.
To make this happen,
in \Cref{sec:prerequisite} \sys{} makes basic assumptions for the deployment framework, 
including a compute graph abstraction and a reference implementation from a mature platform, \ie CUDA here.
To make testing easy, in \Cref{sec:carving}, \sys{} \textit{automatically} creates realistic operator-level tests through test carving, which instruments the execution of each operator on CUDA and collects its input-output pairs as unit tests for the WebGPU support. 
Specifically, \sys{} views the deployment process as a migration task: in the very beginning, we have a reliable CUDA implementation (\ie pure CUDA compute graph) of the model, and we want to eventually run the model fully on WebGPU (\ie pure WebGPU compute graph).
As such, to simplify the mindset,
each time we focus on implementing one operator for WebGPU and validate the operator implementation using the automated tests.
Furthermore, for correctness, 
we also want to make sure that after putting in the newly implemented operator, everything still works as previously.
Therefore, in \Cref{sec:offload}, we maintain an initial CUDA compute graph whose corresponding operators are gradually offloaded to WebGPU once a new WebGPU operator is implemented, and thus, we perform model-wise validation in each step.
The gradual offloading mechanism allows developers to focus on a minimal unit (\ie one operator) each time, which minimizes the debugging scope, and once it passes model-wise validation, we are confident that it will not bring hidden bugs to further steps.
In \Cref{sec:interact}, \sys{} also simplifies these procedures through its universal runtime for flexible cross-platform interaction.
The universal tooling stack in \sys{} can easily run remote computation as if locally and exchange useful information (\eg debugging message) with the target platform.

\subsection{Prerequisites}\label{sec:prerequisite}

While \sys{} is general, it makes the following lightweight assumptions to the deployment process:

\parabf{Compute graph abstraction:}
\sys{} requires the developing system to model ML computation using a compute graph, \ie a directed graph of tensor operators, each of which transforms a set of input tensors to output tensors.
The abstraction of the compute graph allows for straightforward instrumentation of input and output tensors during test carving (in \Cref{sec:carving}).
Viewing an operator as a minimal computation unit also offers a decent granularity for implementation, validation, and offloading to the intended platform (in \Cref{sec:offload}).
This assumption is reasonable as the concept of compute graphs has long existed in various major ML frameworks and compilers, such as TensorFlow~\cite{tf} and PyTorch~\cite{torch}.

\parabf{Reference implementation:}
\sys{} for test automation depends on a \textit{mature} platform as a reference to cross-check the implementation of the target platform.
For example, a mature platform can be the host environment (\eg x86/x64 CPU) or the NVIDIA ecosystem (\eg CUDA).
These platforms have been engineered for decades, achieving the highest level of coverage and reliability.
The assumption is also practical given that CPU and CUDA are the de facto platforms supported by major ML frameworks since the early days.

\subsection{Test Carving}\label{sec:carving}

\newcommand{\opcall}{\mathcal F}
\newcommand{\opinput}{\mathcal I}
\newcommand{\opoutput}{\mathcal O}
\newcommand{\optest}{t}

\begin{figure}
\centering
    \includegraphics[width=0.75\linewidth]{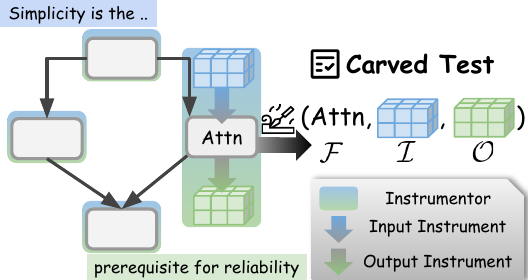}
\caption{Carving intermediate inputs and outputs to a test case.}
\label{fig:carving}
\end{figure}

Test Carving~\cite{carving} is a technique to automatically extract fine-grained unit tests from a given system-level test.
The extracted unit tests are used to replay the intermediate behaviors (\eg individual function calls) that happened during the system test. 
To achieve this,
test carving instruments~\cite{huang1978program} a system test by recording invocations of desired functions and their corresponding inputs and outputs.
By knowing the function, inputs (\eg arguments and context), and outputs (\ie test oracle of differential testing~\cite{mckeeman1998differential}), a test carver synthesizes a unit test to reproduce the expected intermediate step of the system test.

In model deployment, a system test refers to the end-to-end validation of a full model,
where the model is given some semantically meaningful inputs (\eg pictures and sentences) and is expected to produce reasonable end-to-end outputs (\eg correctly completing a motto).
From the high level,
\sys{} carves operator-level unit tests by instrumenting the end-to-end model execution on the source platform, 
to obtain reliable test inputs and reference outputs.
Specifically,
\Cref{fig:carving} illustrates the carving process.
First, the original compute graph on the source platform is modified such that each operator $\opcall_{S}$ is wrapped by an instrumentor, highlighted in the blue-green box.  
The instrumentation includes two instructions:
\emph{(i)} a pre-operator instruction to collect the operator call $\opcall$ and runtime inputs $\opinput$ feed to the instrumented operator;
and 
\emph{(ii)} a post-operator instruction to document the produced output $\opoutput = \opcall_{S}(\opinput)$ as the oracle from the source platform.
Hence, each operator per execution forms a unit test as $\optest = (\opcall, \opinput, \opoutput)$ which can be replayed to validate the target-platform implementation.
Such a source-to-target cross-checking mechanism also indirectly implements Differential Testing~\cite{mckeeman1998differential}.

Notably, the model-wise inputs for bootstrapping the instrumented system test should be semantically meaningful, \eg real-world data from the evaluation test set.
This is important as it allows developers to focus the precision regression over realistic inputs, 
whereas synthetic test inputs, \rebuttal{though meaningful for defending adversarial attacks~\cite{pei2017deepxplore}}, are inadequate to validate the numeric precision stability due to the inconsistent distribution of natural inputs, 
though convenient as it is challenging to specify meaningful high-dimensional embeddings as intermediate inputs.

\subsection{Gradual Target Offloading}\label{sec:offload}

\newcommand{\model}{\mathcal M}
\newcommand{\optestset}{\mathcal T}
\newcommand{\opcallset}{\mathbb F}
\newcommand{\topo}{\mathcal G}
\newcommand{\testmap}{\mathbb T}
\begin{algorithm}
\caption{Gradual Target Offloading Procedure}\label{algo:gradual}
\DontPrintSemicolon
\small
\SetKwProg{Fn}{Function}{:}{}
\SetKw{Continue}{continue}
\SetKw{Break}{break}
\SetKw{Raise}{raise}
\SetKw{Check}{check}
\SetKw{Foreach}{foreach}

\SetKwFunction{GradualTargetOffloading}{\textsc{GradualTargetOffloading}}
\SetKwFunction{Carving}{\textsc{Carving}}
\SetKwFunction{OpWiseValidation}{\textsc{OpWiseValidation}}
\SetKwFunction{ModelWiseValidation}{\textsc{ModelWiseValidation}}

\Fn{\GradualTargetOffloading{$\model_S$}}{
    $\langle \opcallset_S, \topo \rangle \gets \model_S$\;\label{line:decompose}
    $\opcallset_T \gets \{\}$\;\label{line:tdef}
    $\testmap= \{\opcall_i\mapsto\optestset_{i}; i \in 1\cdots \lVert\opcallset_S\rVert \}\gets$\Carving{$\model_S$}\;\label{line:carving}
    \For{$\opcall_{Si} \in \opcallset_S$}{
        $\opcall_{Ti}\gets$ Initial implementation of $\opcall_i$\;\label{line:initimpl}
        $\optestset_{i}\gets \testmap[\opcall_i]$\;
        \While{\OpWiseValidation{$\optestset_{i}$, $\opcall_{Ti}$\label{line:opdebugstart}}}{
            Debugging $\opcall_{Ti}$\;\label{line:opdebugend}
        }
        $\opcallset^*\gets$ $\opcall_{Sk}\mapsto\opcall_{Tk}(\opcallset_S); k \in \arg \left\{\opcallset_T\cup \{\opcall_{Ti}\}\right\}$\;\label{line:substitute}
        \While{\ModelWiseValidation{$\langle\opcallset^*, \topo \rangle$\label{line:mdebugstart}}}{
            Debugging $\opcall_{Ti}$\;\label{line:mdebugend}
        }
        $\opcallset_T\gets \opcallset_T\cup \{\opcall_{Ti}\}$\;\label{line:append}
    }
    $\model_T\gets\langle \opcallset_T, \topo \rangle$\;\label{line:compose}
    \Return $\model_T$\;\label{line:return}
}

\Fn{\OpWiseValidation{$\optestset$, $\opcall$}}{
    \For{$\optest_{i} \in \optestset$}{
        $\langle \opinput, \opoutput \rangle \gets\optest_{i}$\;
        \If{$\opcall(\opinput)\neq\opoutput$}{
            \Return \texttt{False}\;
        }
    }
    \Return \texttt{True}\;
}
\end{algorithm}

\emph{\sys{}} employs the \emph{Gradual Target Offloading} mechanism to progressively transfer the compute graph from a standalone host to a standalone target platform.
\Cref{algo:gradual} details steps of the \emph{Gradual Target Offloading} procedure.
In the high level, \emph{Gradual Target Offloading} takes a source-platform compute graph $\model_S$ as input and migrates it to a target-platform compute graph $\model_T$.
Recall \Cref{sec:prerequisite} that $\model_S$ is a reference implementation from a mature source platform, assumed to be available beforehand.
Both $\model_S$ and $\model_T$ can be initially represented using a general platform-agnostic representation shown in \Cref{fig:codegen}, but they are specialized to different code targets.
Specifically, $\model_S$ can be defined by a list of operators $\opcallset_S$ and their topology information $\topo$ (\Cref{line:decompose}).
Therefore, to obtain $\model_T$ the goal is to transform $\opcallset_S$ to $\opcallset_T$ starting from an empty set (\Cref{line:tdef}).
To start with, we perform test carving (\Cref{sec:carving}) to get a test suite where each operator maps to its corresponding unit tests (\Cref{line:carving}).
For each operator $\opcall_i$ in $\opcallset_S$ we run following migration steps:
\emph{(i)} We draft an initial implementation for $\opcall_i$ on the target platform, \ie $\opcall_{Ti}$ (\Cref{line:initimpl});
\emph{(ii)} we test and debug the implementation of $\opcall_{Ti}$ using the carved tests until the operator-wise validation passes (\Cref{line:opdebugstart,line:opdebugend});
\emph{(iii)} after operator validation, we perform model-wise validation over a hybrid compute graph $\langle\opcallset^*, \topo \rangle$, by substituting corresponding parts in $\opcallset_S$ with implemented target-platform operators, \ie $\left\{\opcallset_T\cup \{\opcall_{Ti}\}\right\}$ (\Cref{line:substitute,line:mdebugstart,line:mdebugend});
and \emph{(iv)} lastly we append the tested $\opcall_{Ti}$ to $\model_T$ to finalize the migration of $\opcall_i$ (\Cref{line:append}).
As such, once all target-platform operators are implemented and tested (\ie $\opcallset_T$), we wrap up the procedure by returning a compute graph $\model_T$ fully supported on the target platform (\Cref{line:compose,line:return}).

\begin{figure}
\centering
    \includegraphics[width=0.75\linewidth]{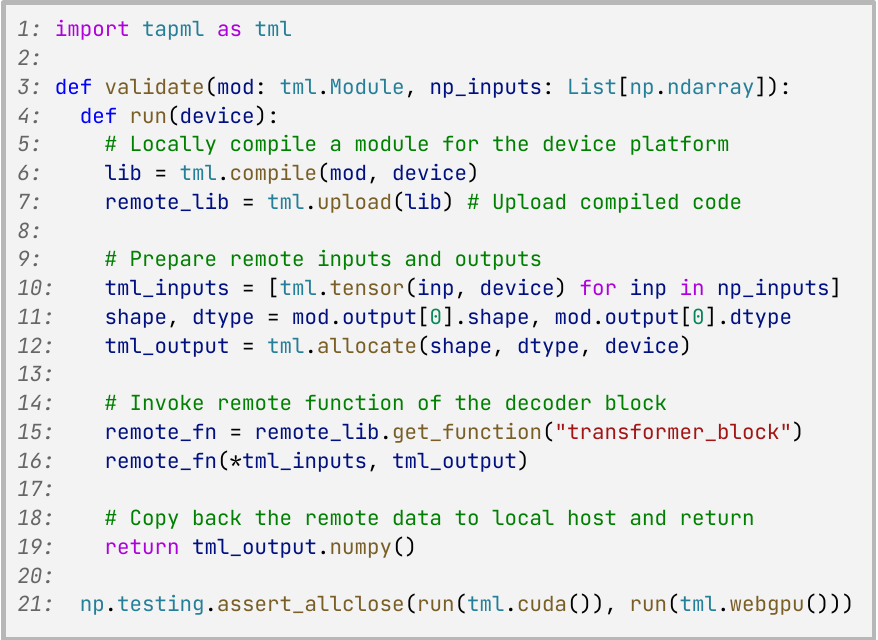}
\caption{Sample validation code using \sys{} APIs.}
\label{fig:uniruntime}
\end{figure}

\subsection{Universal Runtime}\label{sec:interact}

As is shown in \Cref{fig:codegen}, we need a runtime system to transfer data (\eg device code and test inputs) and perform differential testing.
To this end, we build a \emph{universal runtime} to unify these procedures to facilitate cross-platform data transfer and operation.
We argue that it is crucial to unify related behaviors through a standardized interface, which provides a reliable programming model and maximizes code reuse.

\parabf{Exemplification.}
\Cref{fig:uniruntime} demonstrates the interface via an example of differential testing on CUDA and WebGPU implementations. %
Specifically, the tester takes a \texttt{transformer\_block} operator module and a set of computational inputs as the function input (Line~3), based on which we compile the module to both CUDA and WebGPU platforms and perform differential testing (Line~21).
In Line~4, the local function \texttt{run} compiles the \texttt{transformer\_block} module on a given device, returning the results of running the compiled module over the given inputs.
More specifically,
we first compile the target device code locally (Line~6) and upload the compiled device code to the remote device via \texttt{tapml.upload} (Line~7), which returns a handle (\ie reference) to the remote library.
Next, we convert and transfer the test inputs from the local host to the device platform as \texttt{tml\_inputs} in Line~10.
To obtain the output on the device, we also need to allocate space for the output tensor \texttt{tml\_output} in Line~12.
As such, in Line~15 and~16 we call the remote device function, which transforms \texttt{tml\_inputs} to \texttt{tml\_output}, which are transferred back to the local host in Line~19 to perform concrete testing.

\parabf{Implementation.}
The core of the runtime is communication, \ie how to transfer data and instructions across multiple platforms.
Consequently, we developed a lightweight Remote Procedure Call (RPC) framework to manage host-device data transfer and perform device function calling.
Additionally, error handling is a critical yet challenging aspect of universal runtime systems. 
When an error occurs, the runtime is designed to collect error information as much as possible, with complete trackback information, and transmit these debugging messages back to the host before device termination.

\section{Evaluation}

We evaluate \sys{} via the following research questions:

\begin{itemize}
    \item RQ1 (\Cref{sec:eval:support}): How is the overall model support progress by using \sys{}?
    \item RQ2 (\Cref{sec:eval:prod}): How \sys{} improves developer productivity in model debugging?
    \item RQ3 (\Cref{sec:eval:overhead}): What is the overhead of applying \sys{} to model deployments?
    \item RQ4 (\Cref{sec:eval:bugs}): What do the bugs look like when deploying models on emerging platforms?
        How does the \sys{} approach help accelerate the troubleshooting process?
    \item RQ5 (\Cref{sec:eval:practice}): What are the empirical lessons and best practices of using \sys{}?
\end{itemize}

\subsection{Setup}

\parabf{Framework.}
\sys{} is built as the testing and debugging methodology and tooling behind the MLC-LLM~\cite{mlc-llm} project, which focuses on native deployment of \llm{} models on a diverse set of environments across cloud, edge, and emerging platforms like web browsers.
Specifically, as MLC-LLM is built on top of Apache TVM~\cite{tvm}, an open-source ML compiler infrastructure, our implementation of \sys{} relies on the TVM framework to perform various operations, such as manipulating models via its compiler intermediate representation~\cite{relax}.

\begin{table}
\small
    \centering
    \begin{tblr}{row{2,4,6}={bg=azure9}, colspec={Q[c,white] | Q[c] Q[c] Q[c]}}
    \toprule
        & \textbf{Platform}      & \textbf{Hardware}                   & \textbf{RAM Size}  \\
    \midrule
     \SetCell[r=1]{m} Mature
        & CUDA                   & NVIDIA RTX 3080                     & 10 GB \\
    \midrule[dotted]
     \SetCell[r=6]{m} Emerging
        & Vulkan                 & NVIDIA RTX 3080                     & 10 GB \\
        & WebGPU                 & Apple M1 Pro (Macbook Pro)          & 16 GB \\
        & OpenCL                 & Snapdragon 8 Gen 3 (Xiaomi 14 Pro)  & 16 GB \\
        & ROCm                   & Radeon 7900 XTX                     & 24 GB \\
        & \SetCell[r=2]{m}Metal  & Apple M1 Pro (Macbook Pro)          & 16 GB \\ 
        &                        & Apple A18 Pro (iPhone 16 Pro)       & 8 GB\\
    \bottomrule
    \end{tblr}
   \caption{Platforms covered in our experimental setup.}\label{tab:platform}
\end{table}

\parabf{Platforms.}
To demonstrate the versatility and adaptability of \sys{} across a diverse range of configurations,
we evaluate \sys{} over platforms shown in \Cref{tab:platform}.
The corresponding hardware selection spans from a GPU-intensive workstation to energy-efficient mobile devices. 
Furthermore, we consider a rich set of mature and emerging runtime platforms below:

\begin{enumerate}
    \item CUDA~\cite{cuda}: 
        A parallel computing platform and application programming interface model for \textit{NVIDIA GPUs}.
    \item ROCm~\cite{rocm}: 
        An open-source platform by AMD for GPU-accelerated computing, analogous to NVIDIA's CUDA but for \textit{AMD GPUs}. It provides a comprehensive suite of tools and libraries for high-performance, parallel computing applications across various domains.
    \item Metal~\cite{metal}:
        Apple's low-level API for optimizing graphics and parallel computing on \textit{Apple ecosystems}.
        Note that the development setups for iOS-based Metal and macOS-based Metal are quite different;
        yet, we count the two development pipelines as one platform.
    \item WebGPU~\cite{WebGPU}: A web standard providing cross-platform, low-level graphics and computation APIs and capabilities on \textit{web browsers}.
    \item Vulkan~\cite{vulkan}: A \textit{cross-platform} graphics and compute API, providing high-efficiency, low-overhead access to modern GPUs on various devices.
    \item OpenCL~\cite{opencl}: An open and royalty-free standard for \textit{cross-platform} parallel programming of diverse processors found in servers, mobile devices, embedded platforms, etc.
\end{enumerate}

\subsection{RQ1: Platform and Model Support}\label{sec:eval:support}

\begin{table}
\small
    \centering
    \begin{tblr}{row{3,5,7,9}={bg=azure9}, colspec={Q Q[c]  Q[c]}}
    \toprule
    \SetCell[r=2]{c} \textbf{Platform}   & \SetCell[c=2]{m} \textbf{First Available Date} \\
    \cmidrule[lr]{2-3}
                           &  \textbf{Ours} & \textbf{Others} \\
    \midrule
         WebGPU            &  May 1, 2023 & Mar 7, 2024 (Google MediaPipe) \\
         OpenCL on Android &  May 1, 2023 & Mar 7, 2024 (Google MediaPipe) \\
         Metal on iOS      &  May 1, 2023 & Nov 30 2023 (Apple MLX) \\
         Vulkan            &  May 1, 2023 & Sep 18, 2023 (GPT4All) \\
         ROCm              &  May 1, 2023 & Aug 25, 2023 (\texttt{llama.cpp}) \\
         Metal on macOS    &  May 1, 2023 & Jun 5, 2023 (\texttt{llama.cpp}) \\
    \bottomrule
    \end{tblr}
   \caption{Timeline of emerging platform support availability across frameworks.
   Our support is up to 10 months earlier than other open-source and industrial teams. }\label{tab:platformsupp}
\end{table}

\parabf{Overview.}
Starting in April 2023, based on \sys{}, we have deployed \textbf{\numSupportModels{}} models of \textbf{\numSupportArch{}} model architectures over \textbf{\numEmergingPlatforms{}} emerging platforms, empowering emerging AI applications on mobile phones, browsers, servers, and personal laptops.
Meanwhile, to our knowledge,
\Cref{tab:platformsupp} shows that \rebuttal{we are the first in the open-source community to support such a wide range of emerging platforms.}
For example, we supported the mobile and browser platforms 10 months ahead of the MediaPipe engine from Google.
Furthermore, \Cref{tab:modelsup} demonstrates our productivity by showing the gap between the model release date and our full-platform support date.
Our team typically integrates support for new model architectures within 30 days of release. Longer implementation times generally reflect strategic prioritization decisions due to our small team size rather than technical limitations.
It is also worth noting that when we say ``support'', it means the model can not only run on our target platforms but also pass our model-wise testing.

\newcommand{\minicpmfootnote}{\scriptsize MiniCPM authors supported emerging platforms using our framework at their release.}
\newcommand{\llamafootnote}{\scriptsize We started building our framework in April 2023, one month after Llama's release.}
\newcommand{\llavafootnote}{\scriptsize We started supporting vision language models in March 2024.}

\begin{table}
    \centering
\small
    \begin{talltblr}[
        note{*}={\llamafootnote},
        note{$\dag$}={\minicpmfootnote},
        note{$\ddag$}={\llavafootnote},
        caption={Release and support dates across various model architectures and domains.},
        label={tab:modelsup}
    ]{row{2,4,6,8,10,12,14,16}={bg=azure9}, colspec={Q[c,white] | Q[c] | Q[c] Q[c] Q[r]}}
    \toprule
    \textbf{Domain} & \textbf{Arch.} & \textbf{Model Release Date} & \textbf{Our Support Date} & \textbf{Gap Days} \\
    \midrule
    \SetCell[r=11]{} Causal 
    & Llama      & Feb 24, 2023 & Apr 29, 2023~\TblrNote{*} & 64 \\ 
    & Mistral    & Sep 27, 2023 & Sep 28, 2023 & 1 \\
    & Orion      & Jan 20, 2024 & Mar 4, 2024 & 44 \\
    & MiniCPM    & Feb 1, 2024 & Feb 1, 2024~\TblrNote{$\dag$} & 0 \\
    & Qwen2      & Feb 5, 2024 & Feb 13, 2024 & 8 \\
    & StarCoder2 & Feb 28, 2024 & Jul 14, 2024 & 137 \\
    & Qwen2Moe   & Mar 14, 2024 & Apr 5, 2024 & 22 \\
    & Phi3       & Apr 23, 2024 & Apr 25, 2024 & 2 \\
    & Cohere     & May 23, 2024 & Jun 25, 2024 & 33 \\
    & Gemma2     & Jun 27, 2024 & Jul 20, 2024 & 23 \\
    & InternLM2  & Jul 3, 2024 & Jul 7, 2024 & 4 \\
    \midrule[dotted]
    \SetCell[r=1]{} Recursive
    & RWKV6      & Feb 29, 2024 & Mar 14, 2024 & 14 \\
    \midrule[dotted]
    \SetCell[r=2]{} Vision
    & LLaVA      & Oct 6, 2023 & Mar 19, 2024~\TblrNote{$\ddag$} & 165 \\
    & Phi3 V     & May 19, 2024 & Jul 15, 2024 & 57 \\
    \bottomrule
    \end{talltblr}
\end{table}

\newcommand{\rwkvfootnote}{\scriptsize RWKV depends on a special tokenizer, which is currently not supported on web.}
\newcommand{\specfootnote}{\scriptsize
Speculative decoding requires extensive RAM which is not suitable for resource-constrained platforms.
}
\newcommand{\styleplf}[1]{{\footnotesize #1\xspace}}
\newcommand{\narwkv}{{N/A\TblrNote{$\dag$}}}
\newcommand{\naspec}{{N/A\TblrNote{$\ddag$}}}

\begin{table*}[]
    \centering
\small
    \begin{talltblr}[
        caption={\rebuttal{Demonstrating deployment quality by showing the performance of recent models on various backends.}},
        label={tab:modelperf},
        note{$\dag$}={\rwkvfootnote},
        note{$\ddag$}={\specfootnote}
        ]{
        row{5,7,9,11,13,15}={m, bg=azure9},
        colsep={4pt}, rowsep={1pt},
        colspec={@{}Q[c,white, font=\scriptsize] | Q[l,font=\scriptsize] Q[l,font=\scriptsize] | Q[c,font=\scriptsize] Q[c,font=\scriptsize] Q[c,font=\scriptsize] Q[c,font=\scriptsize] Q[c,font=\scriptsize] Q[c,font=\scriptsize] Q[c,font=\scriptsize]@{}},
        }
    \toprule
    & \SetCell[r=3]{m} \textbf{Arch.} & \SetCell[r=3]{c} \textbf{Model} & \SetCell[c=7]{c} \textbf{Tokens / Second} \\
        \cmidrule[lr]{4-10}
        & & & \SetCell[c=2]{c} \textbf{RTX 3080} & & \textbf{7900 XTX} & \SetCell[c=2]{c} \textbf{Apple M1 Pro} & & \textbf{8 Gen 3} & \textbf{A18 Pro}  \\
        \cmidrule[lr,dashed]{4-5} \cmidrule[lr,dashed]{6} \cmidrule[lr,dashed]{7-8} \cmidrule[lr,dashed]{9} \cmidrule[lr,dashed]{10} 
        & & & \styleplf{CUDA} & \styleplf{Vulkan} & \styleplf{ROCm} & \styleplf{Metal} & \styleplf{WebGPU} & \styleplf{OpenCL} & \styleplf{Metal}  \\
    \midrule
    \SetCell[r=11]{} Causal 
    & Llama      & Llama-3.1-8B-Instruct    & 129.5 &  80.1 & 144.2 & 24.7 & 17.0 & 10.2 & 10.2 \\ 
    & Mistral    & Mistral-7B-Instruct      & 141.2 &  83.0 & 156.1 & 24.1 & 19.5 & 10.6 & 11.6 \\
    & Orion      & Orion-14B-Chat           &  74.0 &  41.9 &  85.7 & 13.9 & OOM  & OOM  & OOM  \\
    & MiniCPM    & MiniCPM-2B-128k          & 234.2 & 124.2 & 208.0 & 33.1 & 17.1 & 13.4 & 27.0 \\
    & StarCoder2 & StarCoder2-3B            & 234.5 & 106.6 & 223.2 & 36.7 & 28.5 & 12.9 & 18.6 \\
    & Qwen2Moe   & Qwen1.5-MoE-A2.7B-Chat   & 246.1 & 147.4 & 257.0 & 46.9 & 17.3 & 12.4 & OOM  \\
    & Phi3       & Phi-3.5-mini-instruct    & 226.7 & 142.2 & 239.8 & 32.8 & 32.1 & 11.1 & 16.4 \\
    & Gemma2     & Gemma-2-2B-it            & 279.4 & 139.7 & 247.7 & 39.4 & 21.4 & 12.7 & 28.6 \\
    & InternLM2  & InternLM2.5-1.8B-Chat    & 410.0 & 178.0 & 369.0 & 67.0 & 46.1 & 20.4 & 45.1 \\
    \midrule[dotted]
    \SetCell[r=1]{} Recursive
    & RWKV6      & rwkv-6-world-3b          & 129.0 &  56.6 & 129.5 & 29.5 & {N/A\TblrNote{$\dag$}} & 4.4 & 14.2 \\
    \midrule[dotted]
    \SetCell[r=2]{} Vision
    & LLaVA      & llava-1.5-7b-hf          & 152.7 & 110.4 & 174.4 & 23.1 & 17.5 &  8.2 & 12.5 \\
    & Phi3 V     & Phi-3.5-vision-instruct  & 234.6 & 154.9 & 240.6 & 29.7 & 31.3 & 11.3 & 17.0 \\
    \midrule[dotted]
    \SetCell[r=2]{} {Speculative \\ Decoding}
    & \SetCell[r=2]{} Llama & Llama-3.1-8B-Instruct w/ & \SetCell[r=2]{} 131.0 & \SetCell[r=2]{} 96.2 & \SetCell[r=2]{} 144.5 & \SetCell[r=2]{} 22.9 & \SetCell[r=2]{} \naspec & \SetCell[r=2]{} \naspec & \SetCell[r=2]{} \naspec \\
     & & Llama-3.2-1B-Instruct (Draft) & & & & & & &            \\  
    \bottomrule
    \end{talltblr}
\end{table*}

\parabf{Highlights.}
We exemplify a few of our support highlights below:

\begin{enumerate}
\item \textbf{The first \llm{} on mobile GPUs:}
    Thanks to \sys{}, we are the \emph{first} in open source to successfully deploy \llm{s} on mobile GPUs of Android and iOS devices.
    By maximizing the efficiency of mobile GPUs via Metal and OpenCL, running large \llm{s} on mobile phones becomes a possible and decent experience, \eg running Llama3-8B on Snapdragon 8 Gen 3 at over 10 tokens per second.
    Our pioneering work motivates the community to develop native and mobile \llm{} applications without the fear of sacrificing user privacy.

\item \textbf{The first \llm{} on WebGPU:} 
    \sys{} allows us to be the \emph{first} in public to run \llm{s} on WebGPU, despite its status as an experimental feature in Chrome at the time of development. Enabling emerging ML on WebGPU is of significant broader impact as it allows everyone to run powerful models as easily as opening a web page. 
    Notably, 10 months after our effort, Google introduced the MediaPipe~\cite{mediapipe} via \llm{} Inference API on TensorFlow.js~\cite{tfjs} for WebGPU and mobile platforms.
    We believe \sys{} approach can complement and speed up the development of these related efforts.
\item \textbf{The only \llm{} support on Orange Pi GPU:} 
    To date, we are the \emph{only} team that has successfully deployed \llm{s} on the Orange Pi 5~\cite{orangepi}, an edge device and embedded system in \$100. 
    Yet, we achieve a practical operational speed of 2.5 tokens per second for running Llama2-7B and 5 tokens per second for RedPajama-3B using the OpenCL APIs to maximize the edge GPU on Orange Pi.
    This inspires the community to develop affordable \llm{} applications on edge scenarios.
\end{enumerate}

Thanks to the productivity gain brought by \sys{}, we are able to achieve these highlights and support emerging platforms much earlier than industrial and other open-source teams, despite that we are a team with less than 10 core developers.
Through test carving \sys{} frees developers from writing tedious test cases and ensures the test quality.
Meanwhile, the gradual offloading mechanism in \sys{} minimizes the mental burden of developers by allowing them to focus on the minimal implementation unit of single operators each time until it is fully validated. 
Lastly, the universal runtime becomes a huge time saver since interacting with such emerging and under-developed platforms is made as easy as calling Python functions locally.

\parabf{Deployment quality.}
We demonstrate the quality of our model deployments by showcasing their inference efficiency and benchmark accuracy.
\Cref{tab:modelperf} lists the inference performance of recent models we deployed and quantized in 4 bits.
This result shows that our swift model supports also come with decent quality by being able to serve these models efficiently, even under resource-constrained scenarios.
Beyond efficiency, our deployments have undergone rigorous and productive testing with \sys{}.
\Cref{tab:humaneval} evaluates several recent models deployed by our and other CUDA-focused frameworks using code generation benchmarks,
in which these models excel.
Specifically, we evaluate them using HumanEval+~\cite{liu2024your}, a program synthesis benchmark to rigorously test if \llm{s} can generate functionally correct code.
The results show that our deployed models perform comparably, 
trailing by no more than three tasks relative to the median number of passing tasks achieved by other frameworks' deployments.

Notably, in the \llm{} domain, slight result discrepancy can be expected due to the nature of floating-point arithmetic --- the exact result depends on the floating-point accuracy (\eg 16-bit or even lower precisions are commonly used for \llm{s}) and the detailed computation orders (\eg $a + (b + c) \neq (a + b) + c$) of the implementation.
During decoding, minor differences in decoding probability between two output logits can lead to different, but likely valid, tokens being chosen.
Furthermore, due to the casual nature of \llm{s}, one inconsistent token in the context can lead to entirely different subsequent sentences.

\newcommand{\med}[1]{\textcolor{blue}{$^m$#1}}

\begin{table*}[]
    \centering
\small
    \begin{tblr}{
        row{3,5,7}={m, bg=azure9},
        colspec={lrrrrrrr},
        }
    \toprule
    \SetCell[r=2]{m} Model & \SetCell[r=2]{m} HuggingFace & \SetCell[r=2]{m} vLLM & \SetCell[r=2]{m}SGLang & \SetCell[c=4]{c} Ours \\
        \cmidrule[lr]{5-8}
                           & & & & CUDA & ROCm & Vulkan & Metal \\
    \midrule
 Qwen2.5-Coder-1.5B-Instruct & 101 & \med{107} & 109 & 110 & 110 & 109 & 109 \\
 Llama-3.2-3B-Instruct       &  92 &  98       &  \med{94} &  92 &  92 &  91 &  94 \\
 Gemma-2-2B-it               &  57 &  \med{58} &  61       &  65 &  67 &  65 &  67 \\
 Qwen2.5-0.5B-Instruct       &  42 &  54       &  \med{53} &  52 &  52 &  51 &  51 \\
 
    \bottomrule
    \end{tblr}
    \caption{Number of passing HumanEval+ tasks (164 in total) for recent models deployed by our framework and other CUDA-focused frameworks.
    Our deployments can either outperform (\ie Gemma-2-2B-it) or match the medium of reference results (highlighted as ``\med{$\square$}'') within 3-task discrepancy;
    yet, the popular HuggingFace transformer library can present a notable discrepancy of nearly 10 tasks (\ie Qwen2.5-0.5B-Instruct).}
    \label{tab:humaneval}
\end{table*}

\rebuttal{
\subsection{RQ2: Productivity Improvements}\label{sec:eval:prod}

To evaluate how much \sys{} can improve developer productivity when debugging model deployments,
we designed a controlled experiment using human study to compare \sys{} and bottom-up methods.
In particular, our human study focuses on localizing buggy operators instead of fixing them,
because the efficiency of fixing depends on specific expertise and is orthogonal to the deployment methods.

\parabf{Evaluation setup.}
We recruited three volunteers to debug 10 cases, each of which includes
\emph{(i)} a real-world model including weights and operator information,
\emph{(ii)} a golden CUDA library to run the model, and
\emph{(iii)} a buggy Vulkan library with manually injected bugs in the low-level model IR.
All of the three participants are Ph.D. students in computer science with over 3 to 6 years of experience in ML system engineering.
Each participant is randomly assigned to 3 to 4 cases.
For each case, they are asked to localize the list of problematic operators and we record the debugging time and accuracy of debugging outcomes.
Specifically, for bottom-up debugging, volunteers construct operator-wise test cases to perform differential testing over the two libraries.
They are allowed to access tools that are available in real-world development scenarios, including but not limited to using search engines and \llm{s}, and can create reusable scripts and tools to (partially) automate the workflow.
For top-down debugging, participants can use the \sys{} tool set which can automatically perform gradual target offloading (\Cref{sec:offload}) over all operators.

\parabf{Result overview.}
\Cref{tab:prod} demonstrates that \sys{} brings improvements in not only productivity but also quality in debugging.
Despite the bottom-up baseline on average costing 31 minutes per model to implement tests to identify bugs,
it incurs a false positive rate as high as 58.2\% and a false negative rate of 5.5\%.
In contrast, \sys{} can quickly localize bugs in one minute with no false negatives and a light false positive rate of 10.3\%.
It is also worth noting that case \#1 took significantly longer than the others because it is the first-time setup from the volunteer with the least experience in the toolchain.

\parabf{False positives.}
The bottom-up approach struggled with a high number of false positives (32 in total across all models), with volunteers frequently reporting non-buggy operators.
These false positives are frequent when doing bottom-up debugging for two reasons:
\emph{(i)} models are often quantized for efficient deployment and thus sensitive to the numerical distribution of inputs; and
\emph{(ii)} in bottom-up testing, developers often use random and unrealistic out-of-domain inputs that are not considered by the quantization.
In contrast, the test inputs used in \sys{} are derived from real-world user inputs to meet the input expectation of model quantization, resulting in only 3 false positives from merely 1 model (\ie case \#6).
Nonetheless, \sys{} is not perfect as it also yields 3 false positives in case \#6, which uses a complex Mixture-of-Experts (MoE) model.

\parabf{False negatives.}
The model behind case \#10 is an \llm{} whose Vulkan implementation of paged attention~\cite{vllm} is buggy.
These bugs are missed in bottom-up debugging as our volunteers failed to implement tests for the paged attention kernel.
Implementing paged attention is challenging as it requires a real-world historical key-value (KV) cache with complex page management.
Using random inputs for this kernel usually leads to incompatible configurations and data, crashing the kernel due to API misuse.
On the other hand, \sys{} uses realistic KV caches and configurations derived from the golden library implementation, leading to zero observed false negatives.

Overall, these findings confirm that \sys{'s} top-down design substantially improves the speed and quality of bug localization, providing developers with precise and actionable information about problematic operators,
eliminating the need for extensive manual checking and minimizing the debugging scope.
}

\begin{table}
\small
    \centering
    \begin{tblr}{row{3,5,7,9,11}={bg=azure9}, colspec={Q[c] Q[c] Q[c] Q[r] Q[r] Q[r] Q[r] Q[r] Q[r]  Q[r] Q[r]}}
    \toprule
\SetCell[c=3]{c}\textbf{Model} & & &   
\SetCell[c=4]{c}\textbf{Bottom-up} & & & &
\SetCell[c=4]{c}\textbf{\sys{}} \\
\cmidrule[lr]{1-3}
\cmidrule[lr]{4-7}
\cmidrule[lr]{8-11}
ID   & \# Unique Op. & \# Bugs &
Min. & \# Reports    & \# FP   & \# FN &
Min. & \# Reports    & \# FP   & \# FN        \\
    \bottomrule
1   & 71  & 1      & 74 & 6  & 5 & 0      &  1 & 1  & 0 & 0 \\
2   & 71  & 4      & 22 & 5  & 1 & 0      &  1 & 4  & 0 & 0 \\
3   & 71  & 11     & 13 & 11 & 0 & 0      &  1 & 11 & 0 & 0 \\
4   & 65  & 2      & 29 & 5  & 3 & 0      &  1 & 2  & 0 & 0 \\
5   & 67  & 2      & 21 & 10 & 8 & 0      &  1 & 2  & 0 & 0 \\
6   & 96  & 1      & 71 & 3  & 3 & 1      &  1 & 4  & 3 & 0 \\
7   & 67  & 1      & 18 & 1  & 0 & 0      &  1 & 1  & 0 & 0 \\
8   & 67  & 1      & 16 & 5  & 4 & 0      &  1 & 1  & 0 & 0 \\
9   & 71  & 1      & 20 & 5  & 4 & 0      &  1 & 1  & 0 & 0 \\
10  & 67  & 2      & 28 & 4  & 4 & 2      &  1 & 2  & 0 & 0 \\
    \bottomrule
    \end{tblr}
   \caption{\rebuttal{Debugging productivity and quality comparison between traditional bottom-up approach and \sys{.}}}\label{tab:prod}
\end{table}

\begin{table}
\small
    \centering
    \begin{tblr}{row{2,4,6,8}={bg=azure9}, colspec={Q[c,white] | Q[l] Q[r] Q[c] Q[c]}}
    \toprule
                              & \textbf{Development Step} & \textbf{Platform}  & \textbf{Cost}  & \textbf{One-time Effort?} \\
    \midrule
     \SetCell[r=3]{m} \textbf{Rooting}
                              & Compiling WebGPU model &   Host     & 40.1s & \X \\
                              & Weight loading         &   WebGPU   &  1.6s & \X \\
                              & End-to-end validation  &   WebGPU   &  4.6s & \X \\
    \midrule[dotted]
     \SetCell[r=4]{m} \textbf{Overhead}
                              & Compiling CUDA model   &   Host     & 50.8s & \Y  \\
                              & Weight loading         &   CUDA     & 1.6s  & \Y  \\
                              & Test carving           &   CUDA     & 5.2s  & \Y  \\
                              & Operator validation    &   WebGPU   & 0.1s  & \X  \\
    \bottomrule
    \end{tblr}
   \caption{
   Breakdown \sys{} testing time, exemplified by deploying Llama2-7B on WebGPU with CUDA reference.}\label{tab:overhead}
\end{table}

\subsection{RQ3: Overhead}\label{sec:eval:overhead}

We also systematically examine the testing overhead incurred by applying \sys{} in model deployment, compared to what is necessary.
For clarity, we exemplify the overhead analysis using a case study of deploying Llama2-7B on WebGPU, using CUDA as the reference platform.
The hardware setup is based on a workstation with an AMD Ryzen 9 5900X CPU and an NVIDIA RTX 3080 GPU.
Specifically, Llama2-7B includes 32 decoder blocks; when compiled to the general representation of our base framework,
it leads to 456 operator kernels in 18 operator types.
Consequently, to support Llama2-7B, we need to implement test 18 operator kernel functions and pass $456 \times k$ operator-level tests,
where $k$ is the number of inference tokens determined by the end-to-end testing prompts.

\Cref{tab:overhead} depicts the testing time breakdown of supporting Llama-2 given 6 prefilled token and 1 decoding token.
Specifically, the model testing time is dominated by two parts:

\begin{enumerate}
    \item \emph{Rooting time:} 
        The time for running a model-wise validation on the target platform (\ie WebGPU),
        including compiling the model into WebGPU code (\ie WGSL), initializing the model on WebGPU,
        and running the model on WebGPU with model-wise test inputs.
        These steps are used for end-to-end model validation, which is \textit{indispensable} regardless of the development methods employed.
        Specifically, the bottleneck comes from model compilation where compiling Llama2-7B fully on WebGPU results in 87\% of the end-to-end testing time.
    \item \emph{Overhead time:} 
        Besides routine end-to-end validation, \sys{} additionally requires a full CUDA compute graph for test carving and gradual offloading.
        Compiling Llama2-7B on CUDA, loading its weights, and performing test carving are one-time efforts, which in total can cost less than one minute.
        Furthermore, after implementing each operator, we validate the operator implementation using the carved tests, taking 0.1 seconds on average, which will happen as many times as the number of operator types (\ie 18 times).
\end{enumerate}

To conclude, the overhead introduced by \sys{} is negligible compared to the significant improvement of productivity when developing emerging ML systems.

\subsection{RQ4: Encountered Bugs when Developing Emerging ML Systems}\label{sec:eval:bugs}

\sys{} aims to facilitate the model development on various emerging platforms.
These next-gen platforms are so new that oftentimes they come with limited ecosystem and test coverage, leading to challenging bugs and sophisticated debugging experience. 
Consequently, in addition to our technical contribution, 
we share our concrete experiences of debugging two types of bugs encountered during our development.

\parabf{Bugs from platform drivers.}
The emergence of new computing platforms often brings with it a degree of instability, including inconsistent or incomplete implementations to its specifications. 
For instance, our approach led to the discovery of a \textit{confirmed} concurrency bug in Safari,
a mature and major browser used by billions of Apple users.
We found that one of our WebGPU programs runs flawlessly on Google Chrome but mysteriously produces wrong results on the Technology Preview (Release 188) version of Safari on the same MacBook.
Our diagnosis shows that such discrepancies can be attributed to the incomplete WebGPU API support on Safari, such as \texttt{onSubmittedWorkDone}, whose incompleteness leads to a synchronization bug in which unready buffers were accessed.
This interesting bug finding highlights the effectiveness of \sys{} in capturing challenging and subtle bugs in foundational software stacks before its major release, contributing to developer efficiency and software reliability both at once. 
Furthermore, the base framework by default reuses shared memory buffers of different data types, which is okay for CUDA but introduced Vulkan compilation failures as this feature is not permitted by Vulkan's design.
Additionally, WebGPU is subject to varying constraints across different devices.
For example, when deploying Gemma on WebGPU of the MacOS system, we assumed the maximum number of threads to be the same as that of the native Metal runtime.
However, this leads to \texttt{GPUPipelineError} because WebGPU at most allows for 256 threads on Apple Silicon, which is much smaller than our assumed value.
Similarly, on the PC end, WebGPU allows for 1024MB maximum storage buffer sizes; however, it is limited to 128MB for mobile devices that over-requesting buffer sizes, leading to an error of ``\texttt{Invalid BindGroup}''.

\parabf{Bugs by base framework.}
Our base framework, \ie TVM, is a compilation-based framework that compiles an ML model from its internal representation into target code.
When extending our base framework, a range of issues surfaced, 
predominantly from compiler errors from both the base and extended parts of the framework.
For instance, we observed that our generated Metal code does not perform checks for unaligned data, which leads to incorrect results due to accessing unknown data. 
Meanwhile, missing instruction support in code generation is a major error type discovered by our carved tests.
For example, certain compilation failures were encountered when offloading a matrix multiplication operator to the Metal platform because of unsupported SIMD instructions, such as the selection instructions.
Furthermore, we also encountered performance issues.
For example, our Metal code-generation pipeline initially did not support warp-level instructions, such as \texttt{simd\_shuffle}, and therefore, vanilla instructions are used, leading to sub-optimal performance.
Meanwhile, we also used to assume an inefficient setting of warp sizes on Apple Silicon devices, leading to undesired processing speed.
All of these errors are quickly located, thanks to the test carving and gradual offloading mechanism in \sys{}, which isolates the debugging scope to a single operator each time.

Besides errors from system implementations, the rest of the issues pertain to numerical discrepancies that arise during runtime, making them considerably more challenging to pinpoint. 
For example, we detected numerical errors when running \llm{s} with \texttt{int8} quantization on the Metal platform, yet the same quantization method works perfectly on other platforms.
To isolate this issue, we leveraged automated unit tests carved from the CUDA implementation of the operators and compared the results with those from the Metal platform. 
As such, we quickly identified the root cause in the dequantization kernel, where an unexpected conversion from \texttt{uint8x4} to \texttt{uint32} was erroneously performed prior to converting to \texttt{halfx4}. 
The correct procedure should have been a direct conversion from \texttt{uint8x4} to \texttt{halfx4}. 
Remarkably, this debugging process was completed within just two hours, showcasing the debuggability of \sys{}.

The presented cases empirically demonstrated the huge inconsistency and complexity of various platforms, posing significant challenges in model deployment. 
Nevertheless, the adoption of a top-down methodology has significantly enhanced our ability to support emerging models progressively.
Initially, our adaptation process could span up to a few months based on a traditional bottom-up flow; 
however, with refinements to our approach, we have reduced this time to even a single day, regardless of the additional development complexity involved by the new runtime environments or backend systems. 
This agile response capability underscores the effectiveness of the top-down approach in the dynamic landscape of language model development.

\subsection{RQ5: Lessons and Best Practices}\label{sec:eval:practice}

We present a practical exploration of real-world debugging scenarios, illustrating how the integration of \sys{} with a top-down methodology can be effectively applied to complex, real-life cases. 
We showcase two best practices of using \sys{}, corresponding to our two main sub-techniques.

\parabf{Best practice of test carving:}
Debugging runtime numerical errors, particularly those that occur sporadically, presents a significant challenge. 
Such errors were encountered when developing the Phi-2 model on Metal, leading to the emergence of \texttt{NaN} outputs. %
Interestingly, this problem did not happen on CUDA and was only induced when using certain prompts.

To address this, we employed top-down debugging via \sys{}.
By using our test carvers to automatically break down the CUDA Phi-2 model into individual reliable tests, we automatically examine the correctness of the Metal implementation at a granularity of operators, which allowed us to quickly pinpoint the \texttt{GeLU} operator as the source of erroneous output.
Further investigation revealed that the root cause was the Metal platform's numerically unstable fast-math implementation of the \texttt{tanh} function, which could yield \texttt{NaN} values for inputs exceeding 45. 
We resolved the issue by disabling the fast-math option on the Metal platform, which is replaced by a numerically stable version of the \texttt{tanh} reimplemented by us.

In contrast, a bottom-up strategy for debugging would likely have entailed writing unit tests for each operator, typically with random inputs. 
Such a method, however, would not always trigger the specific error we encountered as it depends on particular activation values that can be induced by real-life input.
Meanwhile, it is also hard to pin down the error in end-to-end integration.
Our experience underscores the importance of a targeted, top-down method to isolate and resolve elusive numerical errors in complex computational models.

\parabf{Best practice of gradual offloading:}
RWKV~\cite{rwkv} is an innovative neural network architecture with RNN architecture and Transformer-level performance. 
Distinct from the prevailing Transformer architectures, RWKV introduces a unique and specialized scan kernel, namely the \texttt{wkv} kernel.
This kernel does not lend itself to straightforward representation as a mere aggregation of conventional operators, which requires support from scratch to deploy on nascent platforms.

The gradual target offloading mechanism in \sys{} facilitates the deployment by allowing us to focus on individual operators.
Initially, we implemented the \texttt{wkv} kernel by describing its IR in our base framework, which is then offloaded to the target platform, with every else model component maintained on the CUDA platform.
As such, we isolate the newly implemented \texttt{wkv} target kernel into the CUDA graph and through end-to-end model validation can fully validate if the new support can tackle real-life inputs.
Thanks to the universal runtime, the detailed steps such as data transfer and remote code execution are all done in a Python environment as if locally, largely lowering the programming barrier.
Consequently, by using \sys{}, we become the first and sole team to run RWKV models across a wide range of platforms.

In contrast, a bottom-up scheme would have necessitated the completion of the entire RWKV model before end-to-end testing on the target platform. 
This could lead to significant challenges in troubleshooting if the eventual integration turns out to be erroneous. 
Furthermore, the \texttt{wkv} kernel's dependency on real-world inputs and the state of preceding tokens complicates the creation of unit tests. 
Given these intricacies, comprehensive end-to-end validation or automatic tests emerge as viable options for progressive model deployment.

\section{Related Work}

\subsection{Emerging ML Models and Platforms}

Machine Learning has demonstrated remarkable capabilities across a variety of tasks. 
Specifically, \llmfull{s}~\cite{gpt4,llama2,mixtral,phi} have shown proficiency in linguistic tasks by accurately predicting subsequent tokens. 
Meanwhile, diffusion models~\cite{imagen,sdxl,dalle3, svd, sora} are adept at producing realistic images and videos in response to textual prompts. 
However, deploying these models presents significant challenges. 
They not only require substantial computational resources but also necessitate repeated executions within a single request.

As these models increase in size and complexity, the desire to deploy them broadly, conveniently, and efficiently persists. 
Major high-performance computing corporations such as NVIDIA, AMD, Intel, and Google have released cutting-edge AI accelerators. 
Concurrently, startups like SambaNova, Cerebras, and Tenstorrent are introducing chips with innovative architectures. 
Additionally, mobile SoCs have incorporated specialized Neural Processing Units (NPUs), \eg Apple's Neural Engine and Qualcomm's Hexagon Tensor Processor.

To close the gap between hardware performance and programmability,
open APIs such as Vulkan and OpenCL enable compatibility with different GPU types, allowing for the construction of universally deployable models on supported hardware without complicated dependency chains. 
More recently, WebGPU unlocks the GPU power within web browsers. 
While these runtimes offer considerable convenience in production, they also pose development challenges due to the nascent state of their associated development and debugging tools.

At the higher level, several recent frameworks have been developed to catch up with the rapid evolution of models.
Examples include vLLM~\cite{vllm}, Hugging Face Transformers~\cite{hf_transformers}, llama.cpp~\cite{llama.cpp}, TensorRT-LLM~\cite{trt-llm}, SGLang~\cite{sglang}, and MediaPipe~\cite{mediapipe}.
However, these frameworks often focus on major or specific platforms, \eg CUDA for NVIDIA GPUs.
\sys{} aims to close the gap by initiating and accelerating the support of underrepresented emerging platforms,
to diversify the possible scenarios of ML applications.

\subsection{Testing and Debugging in ML Systems}

\rebuttal{
Software testing techniques for ML frameworks and compilers have been well studied in recent years to improve their reliability.
These approaches mainly focus on fuzzing-oriented~\cite{miller1990empirical} solutions to develop interesting test cases and extensive test oracles.

For diverse test generation, recent work has been focusing on generating valid and diverse ML models~\cite{freefuzz,xie2022docter,torchprobe,wang2023gencog,deng2023large,nnsmith,neuri} (\eg PyTorch programs) or IRs~\cite{deepdiffer,mlirsmith}.
For example, NNSmith~\cite{nnsmith} applies a solver-aided approach to define operator constraints and generate valid graphs through SMT solving.
\sys{} does not perform model generation since the model to test (\ie the model to be deployed) is determined in model deployment.

Besides test model generation, developing test oracles is also crucial.
The oracle behind \sys{} is differential testing~\cite{mckeeman1998differential,evans2007differential},
which has often served as the de facto oracle in existing ML system fuzzers~\cite{freefuzz,xie2022docter,torchprobe,wang2023gencog,deng2023large,nnsmith,neuri}.
For example, given the same model and model inputs, we run the model computation over different settings in terms of platforms (\eg CPU and GPU), optimization, etc.
More recently, novel differential testing oracles have also been proposed to detect even more bugs.
This includes applying translation validation~\cite{mlir-tv} to test ML compiler passes,
inferring output equivalence via API relationship~\cite{deeprel}, and expanding testing scenarios~\cite{yang2023fuzzing,obsan,wang2022eagle}.

}

Debugging techniques have been used in ML system development to accelerate troubleshooting.
PolyGraphy~\cite{polygraphy} is a tool aiming to diagnose a small error-inducing subgraph out of an ONNX model~\cite{onnx} that fails on NVIDIA TensorRT~\cite{trt}, via test input reduction~\cite{dd}.
Specifically, PolyGraphy either performs bisection or linearly removes the top and bottom layers to get smaller sequence models to reproduce the error.
However, such a test input reduction mechanism is neither optimal nor efficient, as it assumes the input model to be a sequence that can actually be a graph and requires significant attempts of recompilation and testing.
\sys{} addresses the limitations by directly isolating errors in small minimal subgraphs through a migration-based strategy to implement and test one operator each time.

Last, our paper is closely related to software migration~\cite{fleurey2007model}.
However, instead of migrating code~\cite{nguyen2014statistical,zhong2010mining},
\sys{} focuses on platform migration~\cite{behrang2019test,behrang2018automated,qin2019testmig} of ML models.
Such migration requires correct and efficient model implementation using target-platform languages and runtime.
Cross-platform migration is challenging; therefore, the methodology of \sys{} encourages making the process incremental and the result observable per step to minimize the scope of errors.

\section{Conclusion}

\sys{} aims to address the critical software development challenges posed by the rapid evolution of ML models and the emergence of new computing platforms. 
By adopting a top-down approach and utilizing the universal runtime, \sys{} streamlines the deployment process on diverse platforms, significantly enhancing developer productivity. 
This method, characterized by automated testing and a migration-based strategy, has proven effective in our real-world practice,
evidenced by the successful deployment of \numSupportModels{} emerging models across \numEmergingPlatforms{} novel platforms. 
The comprehensive case studies derived from these deployments offer great insights and best practices for developing ML systems on emerging platforms. 
Ultimately, we show that the \sys{} approach accelerates the deployment process while ensuring the quality of emerging models, bridging the gap left by traditional methods and existing frameworks.

\section*{Data Availability}

The artifact of this paper is available at \url{https://zenodo.org/records/15117702}.
The core implementation in \sys{} has been merged and maintained as part of the MLC-LLM project at \url{https://github.com/mlc-ai/mlc-llm}.

\section*{Acknowledgement}

This project would have been impossible without the open-source efforts of the communities behind Apache TVM and MLC-LLM.
Jiawei Liu is supported by a Ph.D. fellowship provided by Amazon.

\bibliographystyle{ACM-Reference-Format}
\bibliography{reference}
\end{document}